\documentclass{amsart}
\usepackage{amssymb} % If your article includes graphics, uncomment this command.
\usepackage{graphicx} % If the article includes commutative diagrams, ... %
\newtheorem{theorem}{Theorem}[section] \newtheorem{lemma}[theorem]{Lemma} \theoremstyle{definition}    \theoremstyle{remark}  \numberwithin{equation}{section}
\theoremstyle{proposition}
\newtheorem{proposition}[theorem]{Proposition}

\begin{document}
\title{Asymptotics of Best-Packing on Rectifiable Sets}
% Only \author and \address are required; other information is % optional. Remove any unused author tags. % author one information %
\author[]{}
\author{S.V. Borodachov} \address{Department of Mathematics, Vanderbilt University, Nashville, TN, 37240} \curraddr{} \email{sergiy.v.borodachov@vanderbilt.edu} \thanks{} % author two information
\author{D.P. Hardin} \address{Department of Mathematics, Vanderbilt University, Nashville, TN, 37240} \curraddr{} \email{doug.hardin@vanderbilt.edu} \thanks{The research of the second author  was supported, in part,
by the U. S. National Science Foundation under grants DMS-0505756 and DMS-0532154.}
\author{E.B. Saff} \address{Department of Mathematics, Vanderbilt University, Nashville, TN, 37240} \curraddr{} \email{Edward.B.Saff@Vanderbilt.edu} \thanks{The research of the third author was supported, in part,
by the U. S. National Science Foundation under grant DMS-0532154.}
% \subjclass is required.
\date{April 17, 2006} %\dedicatory{} % "Communicated by" -- provide editor's name; required.
\commby{} % Abstract is required.
\begin{abstract}We investigate the asymptotic behavior, as $N$ grows,  of the largest minimal pairwise distance of $N$ points restricted to an arbitrary compact rectifiable set
embedded in Euclidean space,    and we find the limit distribution
of such optimal configurations. For this purpose, we compare
best-packing configurations with minimal Riesz $s$-energy
configurations and determine the $s$-th root asymptotic behavior
(as $s\to \infty)$ of the minimal energy constants.

We show that the upper and the lower dimension of a set defined
through the Riesz energy or best-packing coincides with the upper
and lower Minkowski dimension, respectively.

For certain sets in ${\rm {\bf R}}^d$ of integer Hausdorff
dimension, we show that the limiting behavior of the best-packing
distance as well as the minimal $s$-energy for large $s$ is
different for different subsequences of the cardinalities of the
configurations.
 \end{abstract} \maketitle
 % Text of article.

\section {Preliminaries.}

The problem of finding a configuration of $N$ points on the sphere
with the minimal pairwise distance between the points being as
large as possible is classical and is known as {\em Tammes's
problem} or the {\em hard spheres problem}. When formulated for
the whole Euclidean space, the analogous problem is that of
finding a collection (or packing) of non-overlapping equal balls
with the largest density. More information on this problem and its
generalizations can be found in \cite{BorFPC}, \cite{ConSloSPLG},
\cite{LFTRF}, \cite{RogPC}. In the present paper we investigate
the best-packing problem on certain classes of ``non-smooth" sets.

\subsection {Best-packing problem.} We denote by ${\rm {\bf
R}}^{d'}$ the embedding space,  reserving the symbol $d$ for the
dimension of the set being considered. For a collection of $N$
distinct points $\omega_N=\{y_1,\ldots,y_N\}\subset {\rm {\bf
R}}^{d'}$ we set
$$
\delta(\omega_N):=\min_{1\leq i\neq j\leq
N}{\left|y_i-y_j\right|},
$$
and for an infinite set $A\subset {\rm {\bf R}}^{d'}$, we let
\begin{equation}\label {p3} \delta_N(A):=\sup   \{\delta (\omega_N)\
:\ {\omega_N\subset A,\ \#   \omega_N=N}\}
\end{equation} be the {\em best-packing distance} of $N$-point
configurations on $A$, where $\# X$ denotes the cardinality of the
set $X$. We can consider only compact infinite sets $A$, since
quantity (\ref {p3}) is infinite for unbounded sets and is the
same for a set and its closure.

For the case when $A$ is the unit sphere $S^2\subset {\rm {\bf
R}}^3$ exact values of $\delta_N(A)$ have been determined for
$2\leq N\leq 12$ and $N=24$ (see \cite {BorFPC} for references).
For arbitrarily large values of $N$, the precise determination of
best-packing distances is, in general, an intractable problem.
Thus we focus on their asymptotic behavior. For this purpose we
introduce the following notation. Let $0<\alpha\leq d'$ and set
\begin{equation} \label{a1} \underline g_{\infty,\alpha}
(A):=\liminf_{N\to\infty}{\ \delta_N(A)\cdot N^{1/\alpha}},\ \ \ \
\ \overline g_{\infty,\alpha}(A):=\limsup_{N\to\infty}{\
\delta_N(A)\cdot N^{1/\alpha}}. \end{equation} We further put
$$
g_{\infty,\alpha} (A):=\lim_{N\to\infty}{\delta_N(A)\cdot
N^{1/\alpha}},
$$
if this limit exists. On relating these quantities to the {\it
largest sphere packing density} in ${\rm {\bf R}}^d$, $d\in {\rm
{\bf N}}$, which we denote by $\Delta_d$ (see (\ref {density})
below), it can be shown that $g_{\infty,d}([0,1]^d)$ exists and is
given by
\begin{equation} \label {a2}
C_{\infty,d}:=g_{\infty,d}([0,1]^d)=2\left(
{\Delta_d}/{\beta_d}\right)^{1/d}, \end{equation} where $\beta_d$
is the Lebesgue measure (volume) of the unit ball in ${\rm {\bf R}}^d$. It
is not difficult to show that $g_{\infty,d}(A)$ exists for
$d$-dimensional smooth manifolds and domains. Here we shall
establish the existence of $g_{\infty,d}(A)$ for a class of
rectifiable sets and provide a formula for it in terms of the
largest sphere packing density in ${\rm {\bf R}}^d$; we also
describe the limiting distribution of best-packing points (see
Theorem \ref {Th4}).

Recall that the definition of $\Delta_d$ is as follows (cf.
\cite[Chapter 3]{LF-T} or \cite [Chapter 1]{RogPC}). Let $\mathcal
L_d$ stand for the Lebesgue measure in ${\rm {\bf R}}^d$. Denote
by $\Lambda_d$ the set of collections $\mathcal P$ of
non-overlapping unit balls in ${\rm {\bf R}}^d$ for which the
density
$$
\rho (\mathcal P):=\lim_{r\to\infty}{{(2r)^{-d}}}\cdot
\mathcal L_d\left(\cup_{B\in \mathcal P} B\cap
[-r,r]^d\right)
$$
exists. Then \begin{equation}\label {density} \Delta
_d:=\sup\limits_{\mathcal P\in \Lambda _d}{\rho (\mathcal P)}.
\end{equation} It is known that $\Delta_1=1$, $\Delta_2=\pi /\sqrt
{12}$ (Thue in 1892 and Fejes-Toth \cite {LF-T}), and
$\Delta_3=\pi /\sqrt {18}$ (Hales \cite {Hal05}). The exact value
of $\Delta_d$ for $d>3$ is unknown; so far, only upper and lower
estimates, which differ by an exponential factor as $d\to\infty$,
have been obtained for $\Delta _d$ (see \cite {ConSloSPLG} for
this and other references).

Concerning the case when $A=S^2$, the papers \cite {HabvdW51},
\cite{vdW52} prove that $ g_{\infty,2}(S^2)=\left( {8\pi}/{\sqrt
3}\right)^{1/2}. $ Furthermore, the results of \cite {Hal05} imply
that for the unit sphere $S^3\subset {\rm {\bf R}}^4$ we have $
g_{\infty,3}(S^3)=\sqrt 2 \pi^{2/3}. $

\subsection {Minimum energy problem.} The best-packing problem can
be viewed as the limiting case of the problem of minimization of
the discrete Riesz $s$-energy as $s\to\infty$. The setting of this
latter problem is as follows: for $s>0$ and a collection
$\omega_N=\{y_1,\ldots,y_N\}\subset {\rm {\bf R}}^{d'}$, let
$$
E_s(\omega_N):=\sum\limits_{i=1}^{N}{\sum\limits_{j: j\neq
i}{\frac {1}{\left|y_i-y_j\right|^s}}}.
$$
The {\it minimal discrete $N$-point Riesz $s$-energy} of a set $A$
is defined to be \begin{equation}\label {p2} \mathcal
E_s(A,N):=\inf \{E_s(\omega_N) : \omega_N\subset A,\
  \#  \omega_N=N\}. \end{equation}
It is not difficult to see that, for $N$ fixed,
$
\lim_{s\to\infty}{\mathcal E_s(A,N)^{1/s}}={1}/{\delta _N (A)}.
$
We remark that for the case when $s=1$ and $A=S^2$, the
determination of $\mathcal E_s(A,N)$ corresponds to the famous
Thomson problem for the Coulomb potential (cf. \cite {Tho04}).
Exact solutions to this problem are known for $N=2,3,4,6,12$ and
for some interesting cases in higher dimensions (see \cite {Yud93},
\cite {KolYud97}, \cite {And97} and references therein).

When $0<s<n:={\rm dim}_H A$ (the Hausdorff dimension of the set
$A$), potential theory implies that $\mathcal E_s(A,N)\asymp N^2$,
$N\to\infty$ (cf. e.g. \cite{LanFMPT}). When $s>n$, local
interactions dominate, and as described in Theorem 2.1 below, $\mathcal E_s(A,N)\asymp N^{1+s/n}$, $N\to\infty$, for compact rectifiable sets $A$ of positive $n$-dimensional Hausdorff measure.

\subsection {Notation and definitions.} To describe the precise rate of growth of $\mathcal E_s(A,N)$, where $A$ may have non-integer dimension, let $0<\alpha\leq d'$ and for $s>\alpha$ define
\begin{equation} \label{e17} \underline g_{s,\alpha}(A):=\liminf
_{N\to\infty}{\frac {\mathcal E_s(A,N)}{N^{1+s/\alpha}}},\ \ \ \ \
\overline g_{s,\alpha}(A):=\limsup _{N\to\infty}{\frac {\mathcal
E_s(A,N)}{N^{1+s/\alpha}}}
\end{equation} and
\begin {equation}\label{1.7}
g_{s,\alpha}(A):=\lim _{N\to\infty}{\frac {\mathcal
E_s(A,N)}{N^{1+s/\alpha}}},
\end{equation}
if this limit exists. Given a positive integer $d\leq d'$, denote by $\mathcal H_d$
the $d$-dimensional Hausdorff measure in ${\rm {\bf R}}^{d'}$
normalized so that an isometric image of $[0,1]^d$ has measure
$1$. Following \cite {FedGMT} a set $A\subset {\rm {\bf R}}^{d'}$
is called {\it $d$-rectifiable} if it is an image of a bounded
set from ${\rm {\bf R}}^d$ with respect to a Lipschitz mapping. A
set $A\subset {\rm {\bf R}}^{d'}$ is called {\it $\left(\mathcal
H_d,d\right)$-rectifiable}, if $\mathcal H_d(A)<\infty$ and $A$ is
a union of at most a countable collection of $d$-rectifiable sets
and a set of $\mathcal H_d$-measure zero.

Given $\alpha\geq 0$, let $\beta_\alpha=\frac {\pi
^{\alpha/2}}{\Gamma \left(1+\alpha /2 \right)}$. When $\alpha$ is
an integer, this formula agrees with the above definition of
$\beta_d$. Let also $A(\epsilon)$, $\epsilon>0$, be the
$\epsilon$-neighborhood of the set $A\subset {\rm {\bf R}}^{d'}$. {\it The lower and the
upper Minkowski content} of $A$ are defined by
\begin {equation}\label{MnC}
\underline {\mathcal M}_\alpha(A):=\liminf \limits_{\rho \to
0^+}{\frac {\mathcal L_{d'}(A(\rho))}{\beta _{d'-\alpha}\rho
^{d'-\alpha}}}\ \ \ \ {\rm and} \ \ \ \ \overline {\mathcal
M}_\alpha(A):=\limsup \limits_{\rho \to 0^+}{\frac {\mathcal
L_{d'}(A(\rho))}{\beta _{d'-\alpha}\rho ^{d'-\alpha}}},
\end{equation}
respectively. If they coincide, then the quantity $\mathcal
M_\alpha(A):=\underline {\mathcal M}_\alpha(A)=\overline {\mathcal
M}_\alpha(A)$ is called the {\it $\alpha$-dimensional Minkowski
content} of the set $A$. It is known (cf. \cite [Theorem
3.2.39]{FedGMT}) that for every closed $d$-rectifiable set
$A\subset {\rm {\bf R}}^{d'}$ we have
\begin{equation} \label{mc1} \mathcal M_d(A)=\mathcal H_d(A).
\end{equation}

Let $A$ be compact with $\mathcal H_d(A)>0$ and $\{\omega_N\}_{N=2}^{\infty}$ be a
sequence of point configurations on $A$ such that $\#\omega_N=N$,
$N\geq 2$ . We say that $\{\omega_N\}_{N=2}^{\infty}$ is {\it
asymptotically uniformly distributed on $A$ with respect to
$\mathcal H_d$} if, for every subset $B\subset A$ whose boundary
relative to $A$ has $\mathcal H_d$-measure zero, we have
\begin{equation}\label {e15} \frac { \# (\omega_N\cap B)}{N}\to
\frac {\mathcal H_d(B)}{\mathcal H_d(A)},\ \ N\to\infty.
\end{equation}
This definition can also be stated in terms of the weak*
convergence of measures.

\section {Main results.}

First, we describe known results on the asymptotic behavior of the
minimal $s$-energy on sets with ${\rm dim}_H A =d$ and $s>d$. Let
\begin{equation}\label {c_sd} C_{s,d}:=\lim_{N\to\infty}{\frac
{\mathcal E_s\left([0,1]^d,N\right)}{N^{1+s/d}}}, \ \ s>d.
\end{equation} In \cite {MarMayRahSaf04} it is shown that
$C_{s,1}=2\zeta (s)$, $s>1$, where $\zeta
(s)=\sum_{i=1}^{\infty}{i^{-s}}$ is the classical zeta-function.
The articles \cite {HarSaf04}, \cite {HarSaf05} show that for
$s>d$ the limit $C_{s,d}$ exists, is finite and positive. However,
the value of $C_{s,d}$ is still unknown for $d>1$.

The following result is proved in \cite {HarSaf05}, \cite
{BorHarSaf??} (for curves in ${\rm {\bf R}}^{d'}$, it follows from
\cite {MarMayRahSaf04}).
\begin {theorem}\label {rect} Let $s>d$ and $d'\geq d$, where $d$ and $d'$ are integers.
For every infinite compact $(\mathcal H_d,d)$-rectifiable set $A$
in ${\rm {\bf R}}^{d'}$ with $\mathcal M_d(A)=\mathcal H_d(A)$,
the limit $g_{s,d}(A)$ defined in (\ref {1.7}) exists and is given
by
\begin{equation}\label {drect} g_{s,d}(A)={C_{s,d}}{\mathcal
H_d(A)^{-s/d}}.
\end{equation} Moreover, if $A$ is $d$-rectifiable with $\mathcal H_d(A)>0$, then any
sequence $\{\omega^*_N\}_{N=2}^{\infty}$ of $s$-energy minimizing
collections on $A$ such that $  \#   \omega^*_N=N$ is
asymptotically uniformly distributed on $A$ with respect to
$\mathcal H_d$.
\end {theorem}
In the case $d=d'$, Theorem \ref {rect} applies to any compact set $A\subset {\rm {\bf R}}^{d'}$, since such a set is trivially
$d$-rectifiable. In this case $\mathcal H_d=\mathcal L_{d'}$, but we use
the notation $\mathcal H_d$ to handle both the cases $d<d'$ and
$d=d'$.

In this paper we get an analogue of Theorem \ref {rect} for
best-packing configurations.
\begin {theorem}\label {Th4}
Let $d\leq d'$, where $d$, $d'$ are integers, and $A\subset {\rm
{\bf R}}^{d'}$ be an infinite compact $(\mathcal
H_d,d)$-rectifiable set. If $\mathcal M_d(A)=\mathcal H_d(A)$,
then $g_{\infty,d}(A)$ exists and is given by \begin{equation}
\label {e7} g_{\infty,d}(A)=C_{\infty,d}\cdot\mathcal
H_d(A)^{1/d}=2(\Delta _d/\beta_d)^{1/d}\cdot \mathcal
H_d(A)^{1/d}.
\end{equation} Moreover, if $\overline {\mathcal M}_d(A)>\mathcal H_d(A)$, then
\begin{equation}\label {d1} \overline
g_{\infty,d}(A)>C_{\infty,d}\cdot\mathcal H_d(A)^{1/d}.
\end{equation} If $A$ is $d$-rectifiable with $\mathcal H_d(A)>0$,
then every sequence $\{\overline\omega_N\}_{N=2}^{\infty}$ of
best-packing configurations on $A$ such that $\#
\overline\omega_N=N$ is asymptotically uniformly distributed on
$A$ with respect to $\mathcal H_d$.
\end {theorem}
In view of relation (\ref {mc1}), and the fact that any $(\mathcal
H_d,d)$-rectifiable set can be approximated by its $d$-rectifiable
subsets, we either have $\mathcal M_d(A)=\mathcal H_d(A)$ or
$\overline {\mathcal M}_d(A)>\mathcal H_d(A)$, so that either
(\ref {e7}) or (\ref {d1}) must hold.

For results similar to (\ref {e7}) for $d=d'$ that concern the
covering radius, see \cite {GraLus00}.

%In fact, in Theorems \ref {rect} and \ref {Th4} every
%asymptotically optimal sequence of $N$-point configurations on $A$
%will be asymptotically uniformly distributed with respect to the
%$\mathcal H_d$-measure.

We next relate the fundamental constants $C_{s,d}$ and
$C_{\infty,d}$.
\begin {theorem}\label {Th3} The limit
$\lim_{s\to\infty}{C_{s,d}^{1/s}}$ exists for each integer $d>1$ and
$$
\lim\limits_{s\to\infty}{C_{s,d}^{1/s}}=\frac
{1}{C_{\infty,d}}=\frac 12 \left(\frac
{\beta_d}{\Delta_d}\right)^{1/d}.
$$
\end {theorem}

We next show that relation (\ref {drect}) can fail for certain
$(\mathcal H_d,d)$-rectifiable sets.
\begin {proposition}\label {remark1}
If $A\subset {\rm {\bf R}}^{d'}$ is a compact $(\mathcal
H_d,d)$-rectifiable set with $\overline {\mathcal M}_d(A)>\mathcal
H_d(A)$, then for $s$ sufficiently large
$$
\underline g_{s,d}(A)<{C_{s,d}}{\mathcal H_d(A)^{-s/d}}.
$$
\end {proposition}
As an example of a rectifiable set for which (\ref {mc1}) does not
hold, we mention a compact $(\mathcal H_2,2)$-rectifiable set
$B\subset {\rm {\bf R} }^3$ with $0<\mathcal H_2(B)<\infty
=\mathcal M_2(B)$ given in \cite [p. 276]{FedGMT}. Proposition
\ref {MC} will imply that $g_{s,2}(B)=0$, $s>3$, and
$g_{\infty,2}(B)=\infty$.

{\bf Remarks for general sets.} Let $\underline{\rm dim}_MA$ and
$\overline {\rm dim}_MA$ denote the lower and the upper Minkowski
dimension of a set $A\subset {\rm {\bf R}}^{d'}$. One can also
introduce the lower and the upper dimension of a set using
$s$-energy or best-packing. Let $\underline {\rm
dim}_{\infty}A:=\inf(\{\alpha
>0 : \underline g_{\infty, \alpha}(A)=0\}\cup \{d'\})=\sup (\{\alpha \in (0,d'] : \underline g_{\infty, \alpha}(A)=\infty\}\cup \{0\})$ and for
a fixed $s>d'$ denote $\underline {\rm dim}_{s}A:=\inf(\{\alpha >0
: \overline g_{s,\alpha} (A)=\infty\}\cup \{d'\})=\sup (\{\alpha \in
(0,d'] : \overline g_{s,\alpha}(A)=0\}\cup \{0\})$ with $\overline
{\rm dim}_{\infty}A$ and $\overline {\rm dim}_{s}A$ being defined
in an analogous way through $\overline g_{\infty, \alpha}$ or
$\underline g_{s,\alpha}$.
The following proposition implies that for any set  $A\subset {\rm
{\bf R}}^{d'}$ we have $\underline {\rm dim}_{\ \! s}A=\underline
{\rm dim}_{\infty}A=\underline{\rm dim}_MA$ and $\overline {\rm
dim}_{s}A=\overline {\rm dim}_{\infty}A=\overline{\rm dim}_MA$,
provided $s>d'$.

\begin {proposition}\label {MC}
If $\ 0<\alpha \leq d'<s$, there are positive constants
$c_1=c_1(s,\alpha)$ and $c_2=c_2(s,\alpha)$ such that for any
infinite set $A\subset {\rm {\bf R}}^{d'}$ we have
\begin{equation}\label {(1)} c_1\underline{\mathcal M}_\alpha(A)^{-s/\alpha}\leq \overline
g_{s,\alpha}(A)\leq c_2\underline {\mathcal
M}_\alpha(A)^{-s/\alpha},
\end{equation}
\begin{equation} \label {(2)} c_1\overline{\mathcal M}_\alpha(A)^{-s/\alpha}\leq \underline
g_{s,\alpha}(A)\leq c_2\overline {\mathcal
M}_\alpha(A)^{-s/\alpha}.
\end{equation} There are also positive constants $c_3=c_3(\alpha)$ and $c_4=c_4(\alpha)$
such that for every infinite set $A\subset {\rm {\bf R}}^{d'}$
\begin{equation}\label {(3)} c_3\underline {\mathcal M}_\alpha(A)^{1/\alpha}\leq
\underline g_{\infty,\alpha}(A)\leq c_4\underline {\mathcal
M}_\alpha(A)^{1/\alpha},
\end{equation} \begin{equation}\label {(4)} c_3\overline {\mathcal
M}_\alpha(A)^{1/\alpha}\leq \overline g_{\infty,\alpha}(A)\leq
c_4\overline {\mathcal M}_\alpha(A)^{1/\alpha}. \end{equation}
\end {proposition}
It is known that $\underline{\rm dim}_M A\geq {\rm dim}_H A$ with
a strict inequality possible for some compact sets (cf. e.g. \cite
[p. 77]{MatGSMES}). Hence, for such sets $A$ and any ${\rm dim}_H
A<\alpha<\alpha_1<\underline{\rm dim}_M A$ we have $\mathcal
H_\alpha(A)=0$, but $g_{s,\alpha_1}(A)=0$, $s>d'$, and
$g_{\infty,\alpha_1}(A)=\infty$.

For every $s\in (d',\infty]$ and compact sets with sufficiently
large gap between $\underline {\mathcal M}_\alpha (A)$ and
$\overline {\mathcal M}_\alpha (A)$ we will have $\underline
g_{s,\alpha}(A)<\overline g_{s,\alpha }(A)$. Moreover, if
$\underline {\rm dim}_MA<\overline {\rm dim}_MA$ (cf. e.g. \cite
[p. 77]{MatGSMES} for examples), the order of the best-packing
radius and the minimal $s$-energy for $s>d'$ will vary depending
on the subsequence of cardinalities of configurations.

We also show that the condition of $(\mathcal
H_d,d)$-rectifiability in Theorems \ref {rect} and \ref {Th4} is
crucial in the sense that there are non-rectifiable compact sets
with ${\rm dim }_H A=d$ and $0<\mathcal H_d(A)<\infty$ such that
$g_{\infty,d}(A)$ and $g_{s,d}(A)$ (for sufficiently large $s$) do
not exist. Indeed, we show that this is true for a class of
Cantor-type sets which we will denote by $\mathcal K$.

We say that a non-empty compact set $K\subset {\rm {\bf R}}^{d'}$
belongs to the class $\mathcal K$, if there are a finite number of
distinct similitudes $S_1,\ldots,S_p:{\rm {\bf R}}^{d'} \to {\rm
{\bf R}}^{d'}$ with the same contraction coefficient $\sigma \in
(0,1)$ such that \begin{equation}\label {c1} \bigcup
\limits_{i=1}^{p}{S_i(K)}=K, \ \ \ {\rm and}\ \ \ S_i(K)\cap
S_j(K)=\emptyset,\ \ i\neq j. \end{equation} According to \cite
{Hut81}, we have $\lambda:={\rm dim}_H K=-\log_{\sigma}{p}$ and
$0<\mathcal H_\lambda (K)<\infty$. This is a subclass of the class
of self-similar sets constructed in \cite {Hut81} (this
construction is also cited in \cite [Section 4.13]{MatGSMES}).

Class $\mathcal K$ contains the classical Cantor subset of
$[0,1]$. Parameters $p$ and $\sigma$ can also be chosen so that
${\rm dim}_H K$ is any integer between $0$ and $d'$. For example,
if $a_1$, $a_2$ and $a_3$ are vertices of an equilateral triangle
on the plane and $S_i$, $i=1,2,3$, is the homothety of the plane
with respect to $a_i$ and the contraction coefficient $1/3$, then
we get a set of Hausdorff dimension one, known as the Sierpinski gasket \cite [p. 75]{MatGSMES}.

\begin {proposition}\label {Th6}
Let $K$ be a compact set from the class $\mathcal K$ with
$\lambda={\rm dim}_H K$. Then, for $s$ sufficiently large we have
$ 0<\underline g_{s,\lambda}(K)<\overline g_{s,\lambda}(K)<\infty.
$
\end{proposition}

\section {Proofs.}

With regard to the extended real number limits in $[0,\infty]$,
we agree that $1/0=0^{-s}=\infty^{s}=\infty$, $1/\infty=\infty
^{-s}=0$, $s>0$.

\begin {proposition}\label {Th1}
For every infinite set $A\subset {\rm {\bf R}}^{d'}$ and $\
0<\alpha \leq d'$ we have \begin{equation} \label {p1}
\lim_{s\to\infty}\left(\overline
g_{s,\alpha}(A)\right)^{1/s}=\frac {1}{\underline
g_{\infty,\alpha}(A)}\ \ {\rm and}\ \
\lim_{s\to\infty}\left(\underline g_{s,\alpha}(A)\right)^{1/s}=
\frac {1}{\overline g_{\infty,\alpha}(A)}.\end{equation}
\end{proposition}
Proposition \ref {Th1} immediately yields the following
statements.
\begin {proposition}\label {Th2}
Let $A\subset {\rm {\bf R}}^{d'}$ be an infinite set and
$0<\alpha\leq d'$. If for every $s$ sufficiently large $\underline
g_{s,\alpha}(A)=\overline g_{s,\alpha}(A)$, then $g_{\infty,\alpha
}(A) $ exists and
$$\lim_{s\to \infty}\left(g_{s,\alpha}(A)\right)^{1/s}=\frac {1}{g_
{\infty,\alpha} (A)}.$$
\end {proposition}

\begin {proposition}\label {Th5}
Let $A\subset{\rm {\bf R}}^{d'}$ be an infinite set such that
$\underline g_{\infty, \alpha}(A)<\overline g_{\infty,\alpha} (A)$
for some $0<\alpha\leq d'$. Then
%$$
%\lim_{s\to\infty} {\left(\frac {\overline g_{s,\alpha}(A)} {\underline
%g_{s,\alpha}(A)}\right)^{1/s}} =\frac {\overline g_{\infty,\alpha}(A)}
%{\underline g_{\infty,\alpha} (A)}>1
%$$
%and in particular,
for sufficiently large $s$ we have
$
\underline g_{s,\alpha}(A)<\overline g_{s,\alpha}(A).
$
\end {proposition}

{\bf Proof of Proposition \ref {Th1}. Lower estimates.} We can
assume $A\subset {\rm {\bf R}}^{d'}$ to be compact, since on unbounded sets
$g_{s,\alpha} (A)=0$ and $g_{\infty,\alpha}(A)=\infty$ and the
minimal $s$-energy (as well as the best-packing radius) is the
same for $A$ and its closure.

Choose an arbitrary $\epsilon \in (0,1)$ and let $s>\alpha$. Let $N$ be
sufficiently large and $\omega^*_N:=\{x_{1,N},\ldots,x_{N,N}\}$ be
an $s$-energy minimizing $N$-point collection on $A$. Set
$N_\epsilon:=\left\lfloor(1-\epsilon)N\right\rfloor$, where
$\left\lfloor t\right\rfloor$ is the floor function of $t$, and
$$r_{i,N}:=\min_{j:j\neq i}{\left|x_{i,N}-x_{j,N}\right|}.$$ Pick
a point $x_{i_1,N}\in \omega_N^*$ with $r_{i_1,N}\leq
\delta_N(A)$. In $\omega^*_N\setminus \{x_{i_1,N}\}$ pick a point
$x_{i_2,N}$ so that $r_{i_2,N}\leq \delta_{N-1}(A)$. Continue this
process until we pick a point $x_{i_{\lfloor\epsilon
N\rfloor+1},N}\in \omega^*_N\setminus
\{x_{i_1,N},\ldots,x_{i_{\left\lfloor\epsilon
N\right\rfloor},N}\}$ such that $r_{i_{\lfloor\epsilon
N\rfloor+1},N}\leq \delta_{N-\lfloor\epsilon N\rfloor}(A)$. Then
$$
\mathcal E_s(A,N)=E_s(\omega^*_N)\!\geq \!
\sum_{k=1}^{\left\lfloor\epsilon N\right\rfloor+1}\!{\frac
{1}{(r_{{i_k},N})^s}}\!\geq \!\sum_{k=1}^{\left\lfloor\epsilon
N\right\rfloor+1}\!{\frac {1}{(\delta _{N-k+1}(A))^s}}\!\geq
\!\frac {\epsilon N}{\left(\delta _{N_\epsilon}(A)\right)^s}.
$$
Hence,
\begin{equation}\label {p2a} \overline g_{s,\alpha}(A)\geq
\limsup _{N\to\infty}\frac {\epsilon}{\left(\delta
_{N_\epsilon}(A)\right)^sN^{s/\alpha}} =\frac {\epsilon
(1-\epsilon)^{s/\alpha}}{\left(\liminf\limits_{N\to\infty}\delta
_{N_\epsilon}(A)\cdot N_\epsilon ^{1/\alpha}\right)^s}=\frac
{\epsilon (1-\epsilon)^{s/\alpha}}{\left(\underline
g_{\infty,\alpha} (A)\right)^s},
\end{equation} since $N_\epsilon$ passes through all natural
numbers. Similarly,
\begin{equation}\label {3.3}
\underline g_{s,\alpha}(A)\geq \frac {\epsilon
(1-\epsilon)^{s/\alpha}}{\left(\overline g_{\infty,\alpha}
(A)\right)^s}.
\end {equation}
Then, letting first $s\to\infty$ and then $\epsilon \to 0$, we get
\begin{equation}\label
{pd1} \liminf_{s\to\infty}{\left(\overline
g_{s,\alpha}(A)\right)^{1/s}}\geq {\frac {1}{\underline
g_{\infty,\alpha} (A)}}\ \ \ \ {\rm and}\ \ \ \
\liminf_{s\to\infty}{\left(\underline
g_{s,\alpha}(A)\right)^{1/s}}\geq \frac {1}{\overline
g_{\infty,\alpha} (A)}. \end{equation}

{\bf Upper estimates.} Let, for every $N(\geq 2)$ fixed,
$X_N=\{x,x_1,\ldots,x_{N-1}\}\subset {\rm {\bf R}}^{d'}$ be such
that $a:=\delta(X_N)>0$ and for every $k\in {\rm {\bf N}}$ let
$M_k$ be the set of points from $X_N$ contained in $B(x,a(k+1))$
but not in $B(x,ak)$, where $B(x,r)$ is the open ball in ${\rm
{\bf R}}^{d'}$ centered at $x$ with radius $r$. Then, from a volume
argument,
$$
\# M_k\cdot \mathcal
L_{d'}\left[B\left(0,{a/2}\right)\right]\leq \mathcal
L_{d'}\left[ B\left(x,a(k+ 3/2)\right)\setminus
B\left(x,a(k-1/2)\right)\right],
$$
and so $\# M_k\leq (2k+3)^{d'}-(2k-1)^{d'}\leq 4d'(2k+3)^{d'-1}.$
Hence,
$$
\!\!\!\!\!\!\!\!\!\!\!\!\!\!\!\!\!\!\!\!  P_s(x,X_N):=\sum_{i=1}^{N-1}{\frac
{1}{\left|x-x_i\right|^s}}=\sum_{k=1}^{\infty}{\sum_{x_i\in M_{k}
}{\frac {1}{\left|x-x_i\right|^s}}}
$$
$$
\ \ \ \ \ \ \ \ \ \ \ \ \ \ \ \ \ \ \ \ \ \ \leq\sum_{k=1}^{\infty}{\frac {\# M_{k}}{a^sk^s}}\leq \frac
{4d'}{a^s}\sum_{k=1}^{\infty}{\frac {(2k+3)^{d'-1}}{k^s}}
\leq\frac {\eta_{s}}{a^s},\ \ s>d',\nonumber
$$
where $\eta_{s}:=\theta _{d'}\zeta (s-d'+1)$ and $\theta _{d'}$ is
a constant depending only on $d'$.

Now let $\overline \omega_N:=\{\overline x_{1,N},\ldots,\overline
x_{N,N}\}$ be a best-packing $N$-point configuration on $A$; that
is, $\delta (\overline \omega_N)=\delta _N(A)$. Then, using the
above estimate, for $s>d'$ we get
$$
\mathcal E_s(A,N)\leq E_s(\overline
\omega_N)=\sum\limits_{i=1}^{N}{P_s(\overline x_{i,N},\overline
\omega_N)}\leq \frac {\eta_{s}N}{\left(\delta _N(A)\right)^s}.
$$
Hence, for $s>d'$ we have
\begin{equation}\label{pc1}
\overline g_{s,\alpha}(A)\leq \limsup _{N\to\infty}{\frac
{\eta_{s}}{\left(\delta_N(A)\cdot N^{1/\alpha}\right)^s}}=\frac
{\eta_{s}}{\left(\underline g_{\infty,\alpha} (A)\right)^s},\ \ \
\ \underline g_{s,\alpha}(A)\leq \frac {\eta_{s}}{\left(\overline
g_{\infty,\alpha} (A)\right)^s}.\end{equation} Then, since
$\eta_s^{1/s}\to 1$ as $s\to\infty$, we have
\begin{equation}\label{pc2} \limsup _{s\to\infty}{\left(\overline
g_{s,\alpha}(A)\right)^{1/s}}\leq {\frac {1}{\underline
g_{\infty,\alpha} (A)}}, \ \ \ \ \ \ \limsup
_{s\to\infty}{\left(\underline g_{s,\alpha}(A)\right)^{1/s}}\leq
{\frac {1}{\overline g_{\infty,\alpha} (A)}}. \end{equation}
Inequalities (\ref {pd1}) and (\ref {pc2}) yield relations (\ref
{p1}). Propositions \ref {Th1}---\ref {Th5} are proved.

{\bf Proof of Theorem \ref {Th3}.} Using (\ref {c_sd})
and Proposition \ref {Th2} we get
$$
\lim\limits_{s\to\infty}{C_{s,d}^{1/s}}=\lim\limits_{s\to\infty}{g_{s,d}([0,1]^d)^{1/s}}=\frac
{1}{g_{\infty,d}([0,1]^d)}=\frac {1}{C_{\infty,d}}.
$$

{\bf Proof of Theorem \ref {Th4}.} Taking into account Theorem
\ref {rect}, Proposition \ref {Th2} and Theorem \ref {Th3}, we get equation (\ref {e7}):
$$
g_{\infty,d}(A)=\left(\lim\limits_{s\to\infty}{(g_{s,d}(A))^{1/s}}\right)^{-1}=\lim\limits_{s\to\infty}{\frac
{\mathcal H_d(A)^{1/d}}{C_{s,d}^{1/s}}}=C_{\infty,d}\mathcal
H_d(A)^{1/d}.
$$
Now suppose that $A$ is $d$-rectifiable with $\mathcal H_d(A)>0$, and
$\{\overline \omega_N\}_{N=2}^{\infty}$ is a sequence of
best-packing configurations on $A$ such that $ \#
\overline\omega_N=N$. To show that $\{\overline
\omega_N\}_{N=2}^{\infty}$ is asymptotically uniformly distributed
on $A$, choose any subset $B\subset A$ whose boundary relative to
$A$ has $\mathcal H_d$-measure zero. Let $\overline B$ be the
closure of the set $B$.

Set $p_N:=  \#   (\overline \omega_N\cap B)$ and let $\mathcal
N\subset {\rm {\bf N}}$ be an infinite subset such that the limit
$$
p(\mathcal N):=\lim_{\mathcal N\ni N\to\infty}{\frac {p_N}{N}}
$$
exists. If $p(\mathcal N)>0$, then for sufficiently large $N\in
\mathcal N$ we get
$$
\delta_N(A)=\delta(\overline\omega_N)\leq \delta
(\overline\omega_N\cap B)\leq \delta_{p_N}(B)\leq
\delta_{p_N}(\overline B).
$$
Since $\overline B$ is a closed $d$-rectifiable set and $\mathcal
H_d(\overline B)=\mathcal H_d(B)$, using (\ref {e7}), we
have \begin{equation}\label {e10} p(\mathcal N)\leq \lim_{\mathcal
N\ni N\to\infty}{\frac {\delta_{p_N}(\overline B)^d\cdot p_N
}{\delta_N(A)^d\cdot N}}=\left(\frac {g_{\infty,d}(\overline
B)}{g_{\infty, d}(A)}\right)^d=\frac{\mathcal H_d(B)}{\mathcal
H_d(A)}.
\end{equation} If $p(\mathcal N)=0$, then the inequality $p(\mathcal N)\leq
\mathcal H_d(B)/\mathcal H_d(A)$ is trivial. Thus,
$$
\limsup_{N\to\infty} \frac {p_N}{N}\leq {\mathcal
H_d(B)}/{\mathcal H_d(A)}.
$$
Next, let $q_N:=  \# \left(\overline \omega_N\cap (A\setminus
B)\right)$. Since the boundary of $A\setminus B$ relative to $A$
also has $\mathcal H_d$-measure zero, using the same argument we
can write
$$
\limsup_{N\to\infty}{\frac {q_N}{N}}\leq \frac{\mathcal
H_d(A\setminus B)}{\mathcal H_d(A)},
$$
which implies that $ \liminf_{N\to\infty}{p_N}/{N}\geq {\mathcal
H_d(B)}/{\mathcal H_d(A)} $. This shows that (\ref {e15}) holds.

To prove (\ref {d1}) we will need the following lemma. Denote
$\mu_{d'}:=\mathcal L_{d'}(B(0,2))$.
\begin {lemma}\label
{upper} Let $0<\alpha \leq d'$, $G$ and $F$ be two sets in ${\rm
{\bf R}}^{d'}$ and assume that for some positive numbers $c, \gamma$ and $\rho
<(\gamma/\mu_{d'})^{1/\alpha}$ there holds $ \mathcal
L_{d'}\left[G(\rho)\setminus F((c+1)\rho)\right]>\gamma
\rho^{d'-\alpha}. $ Then for $N=\lfloor\gamma
/(\mu_{d'}\rho^\alpha)\rfloor +1$ we have $ \delta_{N}\left(G\setminus
F(c\rho)\right)\geq \rho. $
\end {lemma}
{\bf Proof.} Let $k\in {\rm {\bf N}}\cup\{0\}$ be the largest
number of pairwise disjoint balls of radius $\rho/2$ centered at
points of $G\setminus F (c\rho)$. We just need to show that $ k>
{\gamma}/{(\mu_{d'}\rho^\alpha)}. $ Assume the contrary. Choose points
$x_1,\ldots, x_{k}\in G\setminus F (c\rho)$ such that
$\left|x_i-x_j\right|\geq \rho$, $1\leq i\neq j\leq k$. Then
$$
\mathcal L_{d'}\left(\bigcup
\limits_{i=1}^{k}{B(x_i,2\rho)}\right)\leq k\mu_{d'}\rho^{d'}
\leq\gamma \rho^{d'-\alpha}<\mathcal L_{d'}\left[G(\rho)\setminus
F((c+1)\rho)\right].
$$
This means that there is a point $y\in G(\rho)\setminus
F((c+1)\rho)$ such that $\left|y-x_i\right|\geq 2\rho$,
$i=1,\ldots,k$. Also, there exists a point $x_{k+1}\in G$ such that
$\left|y-x_{k+1}\right|< \rho$. Hence, ${\rm dist
}\left(x_{k+1},F\right)\geq c\rho$. Thus, $x_{k+1}\in G\setminus
F(c\rho)$ and $\left|x_{k+1}-x_i\right|> \rho$, $i=1,\ldots,k$,
and so we have $k+1$ pairwise disjoint balls of radius $\rho/2$
centered at points of $G\setminus F (c\rho)$ which contradicts to
the maximality of $k$. Lemma \ref {upper} is proved.

Another fact needed to show (\ref {d1}) is the left inequality in (\ref {(4)}). We can assume that
$\overline {\mathcal M}_\alpha(A)>0$. Choose any $0<M<\overline
{\mathcal M}_\alpha(A)$. Then there is a sequence
$\{r_m\}_{m=1}^{\infty}$, $r_m\searrow 0$, $m\to\infty$, such that
$\mathcal L_{d'}\left(A(r_m)\right)>M\beta _{d'-\alpha}r_m^{d'-\alpha},\ \ m\in {\rm
{\bf N}}.$ By Lemma \ref {upper} (with $F=\emptyset$) for the sequence  $N_m:=\lfloor M\beta_{d'-\alpha}/(\mu_{d'}r_m^\alpha)\rfloor +1$, $m\in {\rm {\bf N}}$, we have $ \delta_{N_m}(A)\geq r_m\geq 
\left({M\beta_{d'-\alpha}}/{(\mu_{d'} N_m)}\right)^{1/\alpha} $ for sufficiently large
$m$. Hence, $\overline g_{\infty,\alpha} (A)\geq (M\beta_{d'-\alpha}/\mu_{d'})
^{1/\alpha}$. Letting $M\to\overline {\mathcal
M}_\alpha (A)$, gives the lower estimate in (\ref {(4)}).

{\bf Proof of inequality (\ref {d1}).}
%For every $(\mathcal
%H_d,d)$-rectifiable set $A$ we always have $ \mathcal H_d(A)\leq
%\underline {\mathcal M}_d(A), $ and the fact that $\mathcal
%H_d(A)\neq \mathcal M_d(A)$ is equivalent to the inequality
%$\mathcal H_d(A)<\overline {\mathcal M}_d(A)$.
In the case $\mathcal H_d(A)=0$ we have $\overline {\mathcal M}_d(A)>0$ and by the left inequality in (\ref
{(4)}) there holds $\overline g_{\infty,d}(A)>0=C_{\infty,d}\mathcal H_d(A)^{1/d}$. Assume that $\mathcal H_d(A)>0$ and set $d''=d'-d$. Let
$c_0\in (0,1)$ be such that $\left(c_0+1\right)^{d''}\mathcal
H_d(A)<\overline {\mathcal M}_d(A)$ and $M_1,M_2>0$ be such
numbers that
$$
\left(c_0+1\right)^{d''}\mathcal
H_d(A)<\left(c_0+1\right)^{d''}M_1<M_2<\overline {\mathcal
M}_d(A).
$$
Choose any $\epsilon \in (0,1)$ and a $d$-rectifiable compact
subset $K_\epsilon\subset A$ such that $\mathcal
H_d(K_\epsilon)>\mathcal H_d(A)(1-\epsilon)$. By definition (\ref
{MnC}) there is a sequence of positive numbers
$\{r_m\}_{m=1}^{\infty}$, $r_m \searrow 0$, $m\to\infty$, such
that $ \mathcal L_{d'}\left(A(r_m)\right)>M_2\beta_{d''}\cdot
r_m^{d''}, $ $m\in {\rm {\bf N}}$. By (\ref {mc1}) we have $
\mathcal M_d(K_\epsilon)=\mathcal H_d(K_\epsilon)<M_1. $ Then, for
sufficiently large $m$
$$
\mathcal L_{d'}\left[K_\epsilon
((c_0+1)r_m)\right]<M_1\beta_{d''}\cdot(c_0+1)^{d''}r_m^{d''}
$$
and hence,
$$
\mathcal L_{d'}\left[A(r_m)\setminus K_\epsilon
((c_0+1)r_m)\right]>\left(M_2-(c_0+1)^{d''}M_1\right)\beta
_{d''}\cdot r_m^{d''}.
$$
By Lemma \ref {upper} with $\alpha =d$, there is a constant $\nu_1>0$ independent
of $m$ and $\epsilon$, such that for $k_m=\lfloor
\nu_1/r^d_m\rfloor +1$ and $m$ sufficiently large we have
$\delta_{k_m}(A\setminus K_\epsilon (c_0r_m))\geq r_m$. Let $X_m
\subset A\setminus K_\epsilon (c_0r_m)$ be a best-packing
collection of $k_m$ points.

Set $\nu:=C_{\infty,d}\mathcal H_d(A)^{1/d}$. By (\ref {e7})
and the choice of $K_\epsilon$, for sufficiently large $N$, we have
$ \delta_N(K_\epsilon)>{\nu(1-\epsilon)^{1/d}}{N^{-1/d}}. $ Choose
$N_m$ to be the largest integer such that $
{\nu(1-\epsilon)^{1/d}}{N_m^{-1/d}}\geq c_0r_m $ and denote by $Y_m$
the best-packing collection of $N_m$ points on $K_\epsilon$. Since
${\rm dist}(X_m,K_\epsilon)\geq c_0r_m$, we have that
$\delta(X_m\cup Y_m)\geq c_0r_m$ for $m$ sufficiently large.
Hence,
$$
\overline g_{\infty,d}(A)\geq \limsup _{m\to\infty}\delta
_{k_m+N_m}(A)(k_m+N_m)^{1/d}\geq
$$
$$
\geq \limsup _{m\to \infty} {c_0r_m\left(\frac
{\nu_1}{r_m^d}+\frac
{\nu^d(1-\epsilon)}{c_0^dr_m^d}-1\right)^{1/d}}={\left(c_0^d\nu_1+\nu^d(1-\epsilon)\right)^{1/d}}.
$$
Letting $\epsilon \to 0$, we get
$$
\overline g_{\infty,d}(A)\geq
\left(c_0^d\nu_1+\nu^d\right)^{1/d}>\nu=C_{\infty,d}\mathcal
H_d(A)^{1/d}.
$$
This completes the proof of Theorem \ref {Th4}.

{\bf Proof of Proposition \ref {remark1}.} Using Proposition \ref
{Th1},  Theorem \ref {Th3}, and inequality (\ref {d1}) we have
$$
\lim\limits_{s\to\infty}{\left(\frac {\underline
g_{s,d}(A)}{C_{s,d}\mathcal H_{d}(A)^{-s/d}}\right)^{1/s}}=\frac
{C_{\infty,d}\mathcal H_d(A)^{1/d}}{\overline g_{\infty,d}(A)}<1,
$$
and the required inequality follows for sufficiently large $s$.

{\bf Proof of the Proposition \ref {MC}.} We only need to prove
(\ref {(3)}) and (\ref {(4)}) since the upper estimates in (\ref
{(1)}) and (\ref {(2)}) will follow from (\ref {pc1}) and the
lower estimates in (\ref {(3)}) and (\ref {(4)}). Analogously, the
lower estimates in (\ref {(1)}) and (\ref {(2)}) are obtained from
the upper estimates in (\ref {(3)}) and (\ref {(4)}), using (\ref
{p2a}) or (\ref {3.3}) with $\epsilon$ equal, say $1/2$. We remark that (\ref
{p2a}), (\ref {3.3}) and (\ref {pc1}) hold for any infinite set $A$.

Since we do not look for sharp constants, redefine
$$\underline {\mathcal M}_{\alpha}(A):=\liminf _{r\to 0^+}{\mathcal
L_{d'}(A(r))/r^{d'-\alpha}} \ \ \ {\rm and }\ \ \  \overline
{\mathcal M}_{\alpha}(A):=\limsup _{r\to 0^+}{\mathcal
L_{d'}(A(r))/r^{d'-\alpha}}.$$ To show the lower estimate in (\ref
{(3)}), assume that $\underline {\mathcal M}_\alpha (A)>0$
(otherwise it is trivial). Pick any $0<M<\underline {\mathcal
M}_\alpha (A)$ and set $r_N:=(M/(\mu_{d'} N))^{1/\alpha}$. Then, for
$N$ sufficiently large $ \mathcal L_{d'}(A(r_N))>Mr_N^{d'-\alpha}.
$ By Lemma \ref {upper} (with $F=\emptyset$), for
$k_N=\lfloor{M}/{(\mu_{d'} r_N^{\alpha})}\rfloor+1$ ($k_N$ will be
greater that $N$) we have $\delta_{N}(A)\geq \delta_{k_N}(A)\geq
r_N=(M/(\mu_{d'} N))^{1/\alpha}$ for sufficiently large $N$. Hence,
$\underline g_{\infty,\alpha} (A)\geq
\mu_{d'}^{-1/\alpha}M^{1/\alpha}$. Letting $M\to \underline {\mathcal
M}_\alpha (A)$, we get the lower estimate in (\ref {(3)}). We need
the following lemma for the upper estimate.
\begin {lemma}\label {L2}
Let $0<\alpha\leq d'$, $A\neq \emptyset$ be a set in ${\rm {\bf
R}}^{d'}$ and, for some positive numbers $\gamma$ and $\rho
<(\gamma /\beta_{d'})^{1/\alpha}$, assume that there holds $ \mathcal
L_{d'}(A(\rho))< \gamma \rho^{d'-\alpha}. $ Then for any $N>\gamma
/(\beta_{d'}\rho^\alpha)$ we have $\delta_N (A)\leq 2\rho.$
\end {lemma}
{\bf Proof.} Suppose $k\geq 2$ is an integer such that $\delta
_{k}(A)>2\rho$, and let $x_1,\ldots,x_{k}\in A$ be a collection of
distinct points with separation at least $2\rho$. Then
$$
\mathcal
L_{d'}\left(\bigcup\limits_{i=1}^{k}{B(x_i,\rho)}\right)=k\beta_{d'}\rho^{d'}\leq
\mathcal L_{d'}(A(\rho))< \gamma \rho^{d'-\alpha}.
$$
Hence, $k\leq \gamma /(\beta_{d'}\rho^\alpha)$, and so for any $N>
\gamma /(\beta_{d'}\rho^\alpha)$ we have $\delta_N (A)\leq 2\rho$, which proves the lemma.

To get the upper estimate in (\ref {(3)}), we can assume that
$\underline {\mathcal M}_\alpha (A)<\infty$. Choose any $M>
\underline {\mathcal M}_\alpha (A)$. There is a sequence of
positive numbers $\{r_m\}_{m=1}^{\infty}$, $r_m\searrow 0$,
$m\to\infty$, such that $ \mathcal
L_{d'}(A(r_m))<Mr_m^{d'-\alpha},\ \ m\in {\rm {\bf N}}. $ Set
$N_m:=\lfloor{ M/(\beta_{d'}r_m^\alpha)}\rfloor+1$. By Lemma \ref
{L2} we have $\delta_{N_m}(A)\leq 2r_m$ for sufficiently large
$m$. Consequently,
$$
\underline g_{\infty,\alpha} (A)\leq \liminf
_{m\to\infty}{\delta_{N_m}(A)N_m^{1/\alpha}}\leq \liminf
_{m\to\infty}{2r_mN_m^{1/\alpha}}=2\beta _{d'}^{-1/\alpha}
M^{1/\alpha}.
$$
Letting $M\to\underline {\mathcal M}_\alpha (A)$ completes the
proof of (\ref {(3)}).

The left inequality in (\ref {(4)}) was shown before the proof of inequality (\ref {d1}). Thus, it remains to prove the right inequality in (\ref {(4)}) for the
case $\overline {\mathcal M}_\alpha (A)<\infty$. Pick any
$M>\overline {\mathcal M}_\alpha (A)$ and let
$r_N:=(M/(\beta_{d'}(N-1)))^{1/\alpha}$, $N\geq 2$. Then
$
\mathcal L_{d'}\left(A\left(r_N\right)\right)< Mr_N^{d'-\alpha}
$
for $N$ sufficiently large. Since, $N>M/(\beta_{d'}r_N^\alpha)$,
by Lemma \ref {L2} we get $\delta_N (A)\leq 2r_N=
2(M/(\beta_{d'}(N-1)))^{1/\alpha}$. Hence, $\overline
g_{\infty,\alpha}(A)\leq 2\beta _{d'}^{-1/\alpha}M^{1/\alpha}$.
Letting $M\to \overline {\mathcal M}_{\alpha}(A)$ completes the
proof of (\ref {(4)}) and Proposition \ref {MC}.

{\bf Proof of Proposition \ref {Th6}.} It was shown in \cite
{Hut81} (see also \cite [Theorem 4.14]{MatGSMES}) that for any set
$K\in \mathcal K$ there are constants $c_1,c_2>0$ such that
\begin{equation}\label {c_1} c_1r^\lambda\leq \mathcal H_\lambda
\left(K\cap B(x,r)\right)\leq c_2r^\lambda,\ \ \ x\in K,\ \  0<r<
1.
\end{equation} Using an argument analogous to the proof of Lemma \ref {L2}, one can show that $\overline
g_{\infty,\lambda}(K)<\infty$. Since for any set $K\in \mathcal K$
we have $\underline {\mathcal M}_\lambda (K)\geq C\mathcal
H_\lambda (K)>0$ with $C>0$ independent of $K$ (cf. e.g. \cite [p.
79]{MatGSMES}), by (\ref {(3)}) we have $\underline
g_{\infty,\lambda}(K)>0$.

Assume that $g_{\infty,\lambda}(K)$ exists. Let
$S_1,\ldots,S_p:{\rm {\bf R}}^{d'}\to {\rm {\bf R}}^{d'}$ be the
similitudes with the same contraction coefficient $\sigma \in (0,1)$ such that relations (\ref {c1}) hold. Set $h:=\min
_{i\neq j}{\rm dist}\left(S_i(K),S_j(K)\right)$ and choose $k\in
{\rm {\bf N}}$ so that $\delta_k(K)<h$.

Let $m\in {\rm {\bf N}}$ and for ${\rm
{\bf i}}=(i_1,\ldots,i_m)\in \{1,\ldots,p\}^{m}=:Z^m_p$ put
$F_{\rm {\bf i}}:=S_{i_1}\circ \cdots \circ S_{i_m}$. Then
\begin{equation}\label {c3} {\rm dist}\left(F_{\rm
{\bf i}}(K),F_{\rm {\bf j}}(K)\right)\geq h\sigma
^{m-1}>\sigma^{m-1}\delta_k(K),\ \ {\rm {\bf i}}\neq {\rm {\bf
j}},
\end{equation} and $ \bigcup_{{\rm
{\bf i}}\in Z^m_p}{F_{\rm {\bf i}}(K)=K}. $ Let $\overline\omega
_k\subset K$ be a collection of $k$ points such that $\delta
(\overline\omega_k)=\delta_k(K)$, and $ \omega_m:=\cup_{{\rm {\bf
i}}\in Z^m_p}F_{\rm {\bf i}}\left(\overline \omega_k\right). $ In
view of (\ref {c3}), it is not difficult to see that $\delta
_{kp^m}(K)\geq \delta (\omega_m)= \sigma^m\delta _k(K)$. On the
other hand, from any collection of $c_m:=(k-1)p^m+1$ points on $K$
at least $k$ must belong to the same $F_{\rm {\bf i}}(K)$, and
hence, $\delta _{c_m}(K)\leq \sigma^m \delta _k(K)$. Since
$c_m\leq kp^m$, we have $\delta _{kp^m} (K)=\delta _{c_m}(K)$ and
$$
g_{\infty,\lambda}(K)=\lim_{m\to\infty}{\delta_{kp^m}(K)\left(kp^m\right)^{1/\lambda}}=\lim_{m\to\infty}{\delta_{c_m}(K)\left(kp^m\right)^{1/\lambda}}=
$$
$$
=g_{\infty,\lambda}(K)\lim_{m\to\infty}{\left(
{kp^m}/{c_m}\right)^{1/\lambda}}=g_{\infty,\lambda}(K)\left(
{k}/{(k-1)}\right)^{1/\lambda},
$$
which contradicts the finiteness and positiveness of
$g_{\infty,\lambda}(K)$. Hence, $0<\underline
g_{\infty,\lambda}(K)<\overline g_{\infty,\lambda}(K)<\infty$.
Taking into account Propositions \ref {Th1} and \ref {Th5}, we get
Proposition \ref {Th6}.

 % Bibliographies can be prepared with BibTeX using amsplain, % amsalpha, or (for "historical" overviews) natbib style.
 
\end{document}